\documentclass{iau} 
\usepackage{graphicx}
\usepackage{natbib}
\usepackage{amsmath}
\usepackage{threeparttable}

\bibpunct{(}{)}{;}{a}{}{,}

\title[IAUS 342 X-ray observations of Perseus $\&$ clusters]
{Thermal and non-thermal connection\\ in radio mini-halos}

\author[A. Ignesti et al.]   
{A. Ignesti$^{1,2}$, G. Brunetti$^2$, M. Gitti$^{1,2}$, S. Giacintucci$^3$}
 
\affiliation{$^1$DIFA, University of Bologna, Via Gobetti 93/2, 40129 Bologna, Italy \\ email: {\tt alessandro.ignesti2@unibo.it} \\
$^2$IRA INAF, Via Gobetti 101, 40129 Bologna, Italy\\
$^3$U.S. NRL, 4555 Overlook Avenue SW, Code 7213, Washington, DC 20375, USA}

\pubyear{2018}
\volume{342}  
\jname{Perseus in Sicily: from black hole to cluster outskirts}
\editors{K. Asada, E. de Gouveia dal Pino, H. Nagai,\\ R. Nemmen \& M. Giroletti, eds.}
\begin{document}

\maketitle

\begin{abstract}
Several cool-core clusters are known to host a radio mini-halo, a diffuse, steep-spectrum radio source located in their cores, thus probing the presence of non-thermal components as magnetic field and relativistic particles on scales not directly influenced by the central AGN. The nature of the mechanism that produces a population of radio-emitting relativistic particles on the scale of hundreds of kiloparsecs is still unclear.  At the same time, it is still debated if the central AGN may play a role in the formation of mini-halos by providing the seed of the relativistic particles. 
We aim to investigate these open issues by studying the connection between thermal and non-thermal components of the intra-cluster medium. 
We performed a point-to-point analysis of the radio and the X-ray surface brightness of a compilation of mini-halos.
We find that mini-halos have super-linear scalings between radio and X-rays, with radio brightness declining more steeply than the X-ray brightness. This trend is opposite to that generally observed in giant radio halos, thus marking a possible difference in the physics of the two radio sources.
Finally, using the scalings between radio and X-rays and assuming a hadronic origin of mini-halos we derive constraints on the magnetic field in the core of the hosting clusters.
\keywords{radiation mechanisms: nonthermal, radiation mechanisms: thermal, galaxies: clusters: general}
\end{abstract}

\firstsection 
\vspace{-0.5cm}
\section{Introduction}
 About 70$\%$ of relaxed galaxy clusters show cool cores with peaked X-ray surface brightness where the intra-cluster medium (ICM) is efficiently radiating away its thermal energy \citep[e.g.,][]{Hudson_2010}. Those clusters are characterized by the presence of luminous active galactic nuclei (AGN) at their centre \citep[e.g.,][]{Sun_2009b} and magnetic field of the order of ~10 $\mu$G \citep[e.g.,][]{Carilli_2002}. Most of them also show diffuse radio emission in their centre in form of a radio mini-halos (MHs), diffuse radio sources with a radius $R_\text{MH}<0.2R_{500}\simeq 100$ kpc \citep[e.g.,][]{Giacintucci_2017} and steep spectral index ($\alpha>1$ with $S\propto \nu^{-\alpha}$). The main open problem regarding the origin of MHs is the slow-diffusion problem, which arises from the fact that the diffusion time required to the cosmic rays electrons (CRe) to propagate up to the scale of the radio emission is longer ($>10^{9}$ yr) than their radiative time ($\leq 10^{8}$ yr) \citep[e.g.,][]{Brunetti-Jones_2014}. 
There are two possible solutions to this problem. One is the leptonic scenario, where the CRe injected by the central AGN are re-accelerated by ICM turbulence \cite[e.g.,][]{Gitti_2002}. The other is the hadronic scenario, where CRe are produced by collisions between cosmic ray protons (CRp) and thermal protons \citep[e.g.,][]{Pfrommer-Ensslin_2004}. Contrary to the case of giant radio halos, current $\gamma$-ray observations do not provide stringent constraints to the hadronic scenario \citep[e.g.,][]{Brunetti-Jones_2014,Ahnen_2016}. In both models the central AGN may play an important role in the origin of MHs as source of CRs.
\vspace{-0.5cm}
\section{Thermal and non-thermal connection}
Radio MHs represent valuable probes to investigate the properties of cool cores, because they link the thermal and non-thermal components of the ICM. On the one hand, the X-ray emission of the ICM is produced via thermal bremmstrahlung, whose bolometric emissivity is $j_\text{X}\propto n_\text{th}^2T^{1/2}$, where $n_\text{th}$ and $T$ are the particle  density and the temperature of the ICM. On the other hand, the radio synchrotron emissivity depends on the magnetic field and on the energy distribution of CRe. For the hadronic model the injected energy rate of CRe depends on the densities of CRp and thermal protons \citep[e.g.,][]{Pfrommer-Ensslin_2004}, whereas for the leptonic model the CRe turbulent heating depends on the turbulent energy flux and turbulent dissipation scales \citep[e.g.,][]{Brunetti-Lazarian_2016}. Thus, both the radio and X-ray emissivities depend on the density of the thermal plasma, inducing possible scalings between the radio surface brightness, $I_\text{R}$, and the X-ray surface brightness, $I_\text{X}$, in the form:
  \begin{equation}
 I_\text{R}\propto I_\text{X}^{k}
\end{equation}
where the index $k$ can be used to study the non-thermal components.\\

\citet[][]{Feretti_2001} and \citet[][]{Govoni_2001} analyzed the scalings between radio and X-ray brightness in giant radio halos by looking at the point-to-point ({\it ptp}) correlations. They analyzed five radio halos, finding a sub-linear scaling ($k\leq1$).
In our work we want to estimate the $I_\text{R}$-$I_\text{X}$ scaling for a sample of MHs to study the ICM properties in the cool core. 
\section{Sample selection and Monte Carlo point-to-point analysis}
We select our sample of MHs by collecting from literature radio maps with a resolution suitable for the {\it ptp} analysis. For these clusters we then reduced archival Chandra observations to obtain the $I_\text{X}$ images (0.5-2.0 keV). We report in Tab. 1 the properties of our sample and the frequencies of the radio maps that we used.\\
\begin{table}
\begin{center}
\caption{ Physical properties of the sample and results of the Monte Carlo {\it ptp} analysis.}
\begin{tabular}{lcccccccc}
\hline
Cluster name&ID&RA$_\text{J2000}$&DEC$_\text{J2000}$&$z$&R$_{500}^{\dag}$&M$_{500}^{\dag}$&Radio freq.&$k_\text{MC}^{\ddag}$\\
&&[h,m,s]&[deg,',"]&&[Mpc]&[$10^{14}$ $M_{\odot}$]&[GHz]\\
\hline
2A0335&S1(a,b)&03 38 44.4&+09 56 34 &0.035&0.92&2.3$^{+0.2}_{-0.3}$&1.4, 5.5&1.2, 1.1\\
RBS 797&S2&09 47 00.2&+76 23 44&0.345&1.16&6.3$^{+0.6}_{-0.7}$&1.4&1.3\\
Abell 3444&S3(a,b)&10 23 54.8&-27 17 09&0.254&1.27&7.6$^{+0.5}_{-0.6}$&0.6, 1.4&1.2, 1.2\\
MS 1455&S4&14 57 15.1&+22 20 34&0.258&0.98&3.5$^{+0.4}_{-0.4}$&0.6&0.9\\
RXC J1504&S5&15 04 05.4&-02 47 54&0.215&0.98&7.0$^{+0.6}_{-0.6}$&0.3&2.0\\
RX J1532&S6&15 32 53.8&+30 20 58&0.345&1.04&4.7$^{+0.6}_{-0.6}$&1.4&1.1\\
RX J1720&S7&17 20 12.6&+26 37 23&0.164&1.24&6.3$^{+0.4}_{-0.4}$&0.6&1.2\\
\hline
\label{phys.tab}
\end{tabular}
\end{center}
\begin{tablenotes}
\item \textbf{Notes.} $^{\dag}$ Radius and total mass at a mean overdensity of 500 with respect to the cosmological critical density at redshift $z$ \citep[e.g.,][]{Giacintucci_2017}; $^{\ddag}$ The dispersions are $\leq 10 \%$.
\end{tablenotes}
\end{table}
The {\it ptp} analysis is a straightforward method to estimate the $I_\text{R}$-$I_\text{X}$ connection by fitting the values of $I_\text{R}$ and $I_\text{X}$ that are measured in the cells of a mesh tailored on the radio diffuse emission (see Fig. 1). However, different {\it ptp} analysis performed with different meshes may produce different estimates of $k$ for the same MH. 
In order to properly account for this bias related to the arbitrary choice of a particular mesh, we introduce the Monte Carlo {\it ptp} analysis that relies on repeated random sampling of the diffuse radio and X-ray emission. Basically, we perform 1000 cycles of {\it ptp} analysis with a randomly-generated mesh for each cycle. Each mesh is modeled to sample only the radio emission over a given $I_\text{R}$ threshold (we used the 3$\sigma$ level), while excluding the emission not related to the MH (AGN, radio-filled cavities or field sources), and produces a different estimate of $k$ ($k_\text{SM}$). Therefore, we end up with a distribution of values of $k$, instead of a single estimate, and our final estimate for the scaling, $k_\text{MC}$, is given by the mean of this distribution.
\begin{figure}
\minipage{0.47\textwidth}
  \includegraphics[width=\linewidth]{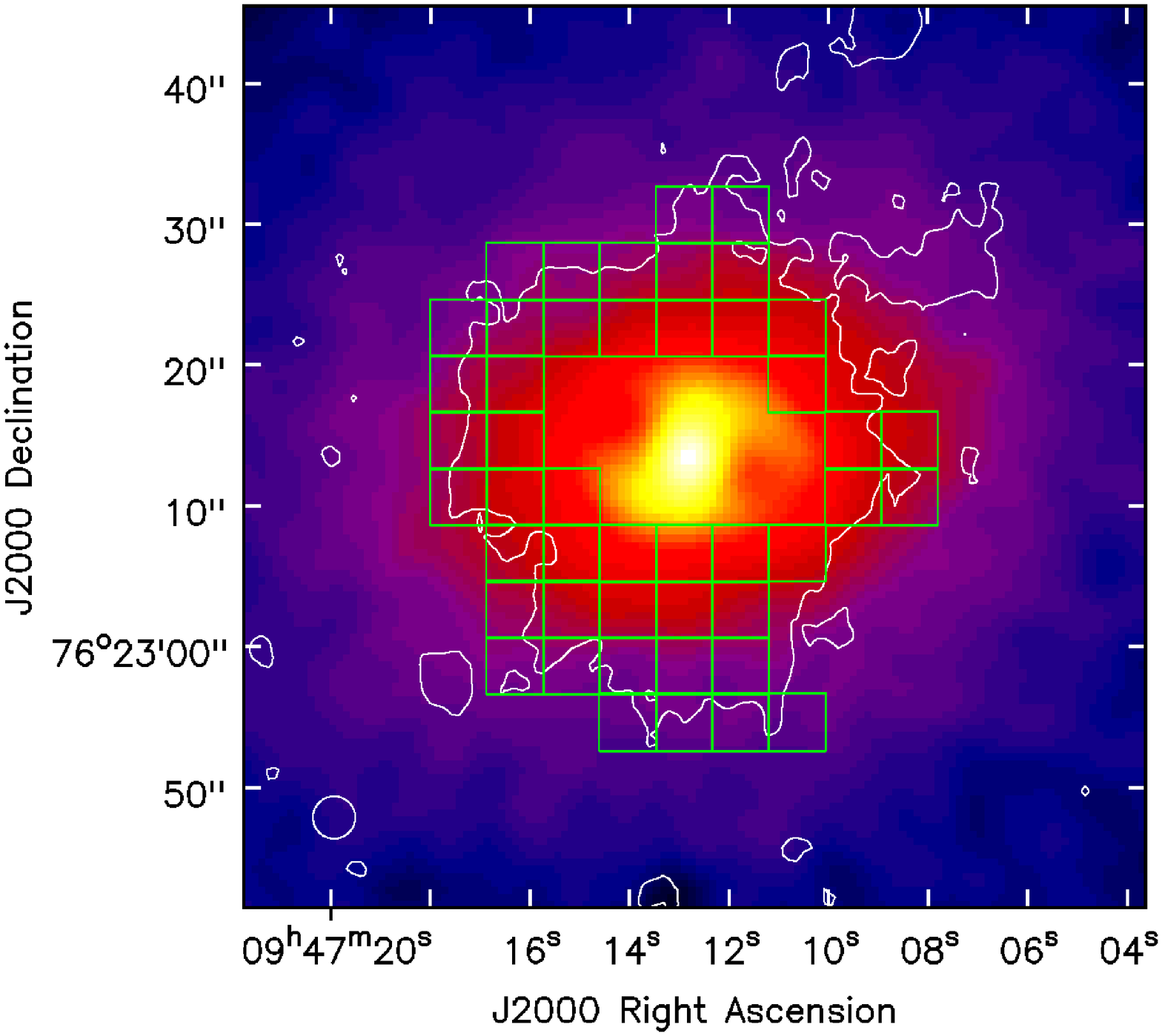}
\endminipage\hfill
\minipage{0.47\textwidth}
  \includegraphics[width=\linewidth]{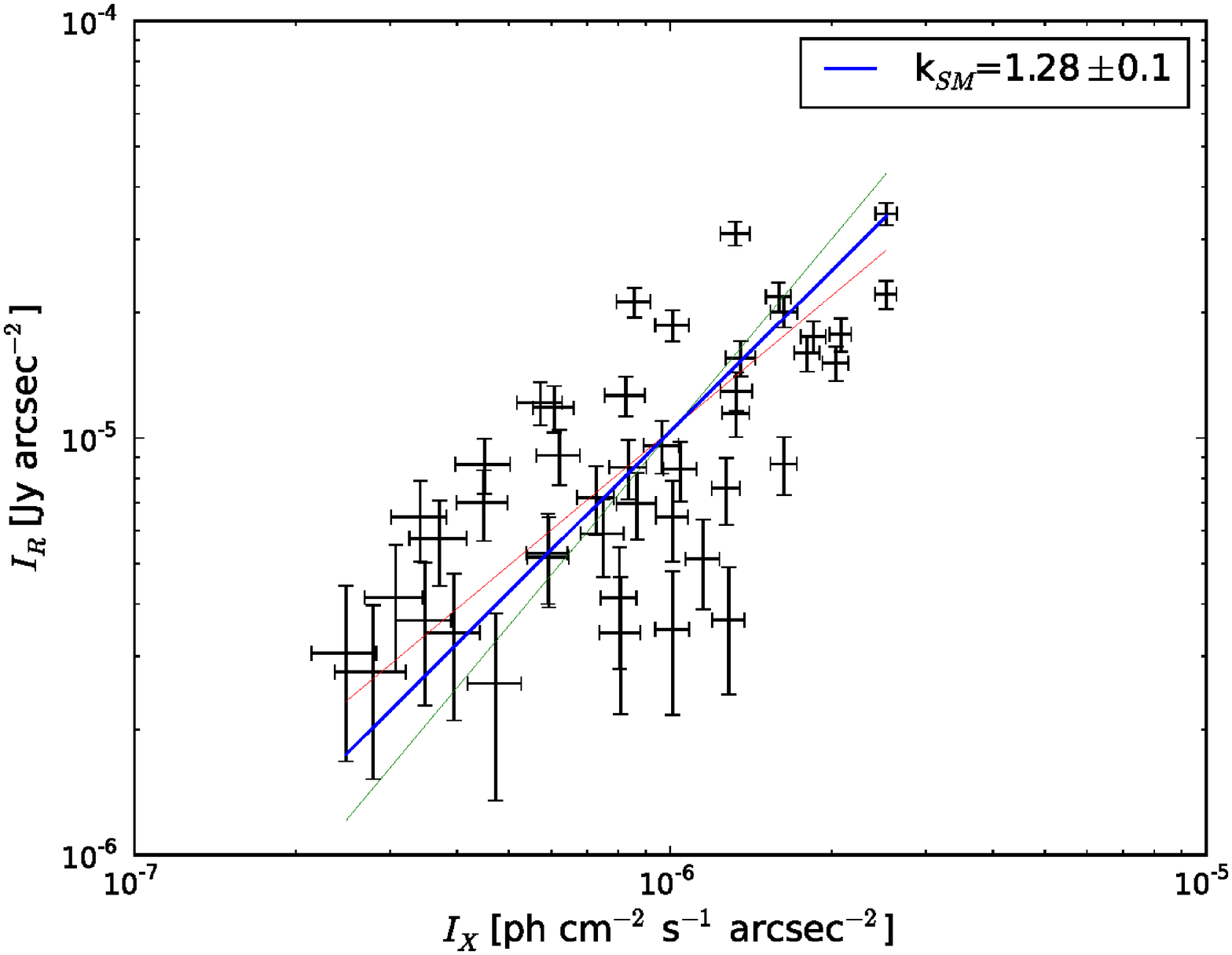}
\endminipage\hfill
\caption{{\it Left:} Chandra 0.5-2.0 keV X-ray image of RBS 797 with the 3$\sigma$ contour (white) of the 1.4 GHz radio emission \citep[1$\sigma$=10 $\mu$Jy beam$^{-1}$,][]{Doria_2012} and a sampling mesh (green); {\it Right:} $I_\text{R}$ vs $I_\text{X}$ measured in the cells of the mesh. The index of the best-fit power-law (blue) is reported in the legend.}
\end{figure}
\vspace{-0.5cm}
\section{Results and implications}
The results of the Monte Carlo {\it ptp} analysis are reported in Tab. 1. In Fig. 1 (right panel) we show the scaling observed in RBS 797 as an example.  
We observe that MHs have super-linear $I_\text{R}$-$I_\text{X}$ scaling ($k>1$). This is opposite to what was observed for giant radio halos ($k\leq$1), thus this difference may indicate an intrinsic physical difference between these objects (Fig. 2, left panel). \\
The $I_\text{R}$-$I_\text{X}$ scaling may allow us to constrain the intra-cluster magnetic field. We assumed an hadronic model for the origin of CRe, in this case the radio emissivity is:
\begin{equation}
 j_\text{R}\propto \epsilon_\text{CRp}n_\text{th}\frac{B^{\alpha+1}}{B^{2}+B_\text{CMB}^{2}}
 \end{equation}
where $B_\text{CMB}=3.25(1+z)^{2}$ $\mu$G is the equivalent magnetic field of the cosmic microwave background and $ \epsilon_\text{CRp}$ is the CRp energy density. In our model we assume that these CRp are continuously injected by the central AGN and diffuse in the core of the cluster. Under stationary conditions, it is $ \epsilon_\text{CRp}\propto Q_\text{p}/r$ where $r$ is the distance from the central source and $Q_\text{p}$ is the CRp injected luminosity \citep[][]{Blasi-Colafrancesco_1999}. We assume the isothermal $\beta$-model \citep[][]{Cavaliere-Fusco_1976} to describe the radial distribution of $n_\text{th}$ inside the cool core and we parametrize the intra-cluster magnetic field as:
\begin{equation}
 B(r)=B_{0}\left[\frac{n_\text{th}(r)}{n_0}\right]^{\eta}
\end{equation}
where $B_0$ and $n_0$ are the central values of the magnetic field and the number density of thermal particles and $\eta$ describes the scaling between the ICM density and the magnetic field.
For the X-ray emissivity we assume $ j_\text{X}\propto n_\text{th}^{2} $, that is the thermal bremmstrahlung emissivity in the case of isothermal core.\\

Therefore, the scaling $I_\text{R}$-$I_\text{X}$ may allow us to get insights into the magnetic field of the ICM and the CR luminosity of the AGN by constraining the free parameters of our model $B_{0}$, $\eta$ and $Q_\text{p}$.
We integrated numerically the emissivity models to obtain a theoretical estimate of $k$ as function of $B_{0}$ and $\eta$ that we compared with the observed scaling. We tested the equipartion ($\eta=1/2$), the isotropically compressed \citep[$\eta=2/3$,][]{Tribble_1993} and the uniform ($\eta=0$) scalings for the magnetic field. Here we report the preliminary results for RBS 797 in Fig.2, right panel. We found that the observed $I_\text{R}$-$I_\text{X}$ scaling is reproduced with $B_{0}\simeq20$ $\mu G$ for $\eta=1/2$ and $B_{0}\simeq50$ $\mu G$ for $\eta=2/3$, whereas the uniform magnetic field can not generate a super-linear scaling. Results for the whole sample and the calculation of the $\gamma$-ray expectations will be presented in Ignesti et al., in preparation.\\
\begin{figure}
\minipage{0.55\textwidth}
  \includegraphics[width=\linewidth]{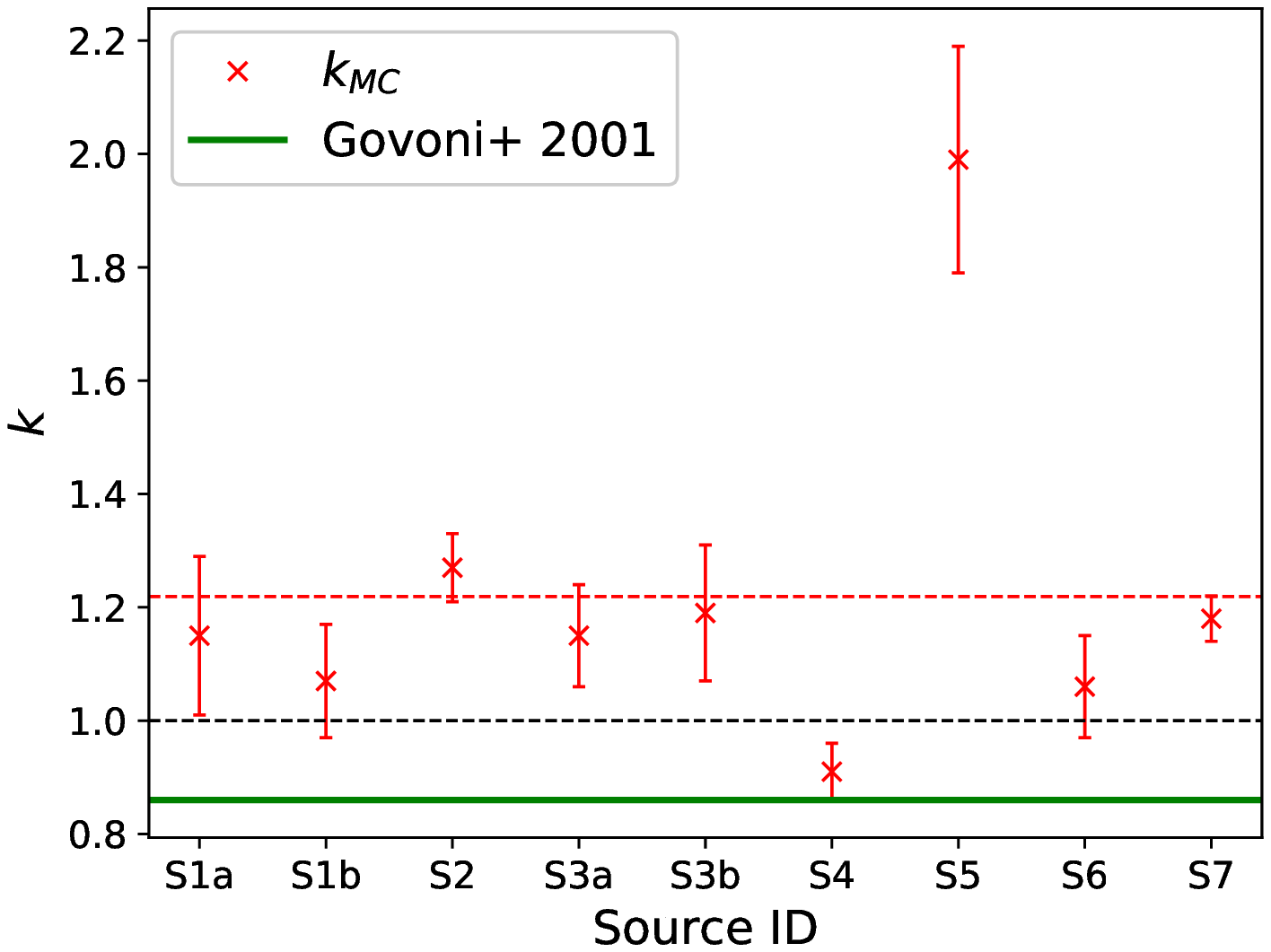}
\endminipage\hfill
\minipage{0.45\textwidth}
  \includegraphics[width=\linewidth]{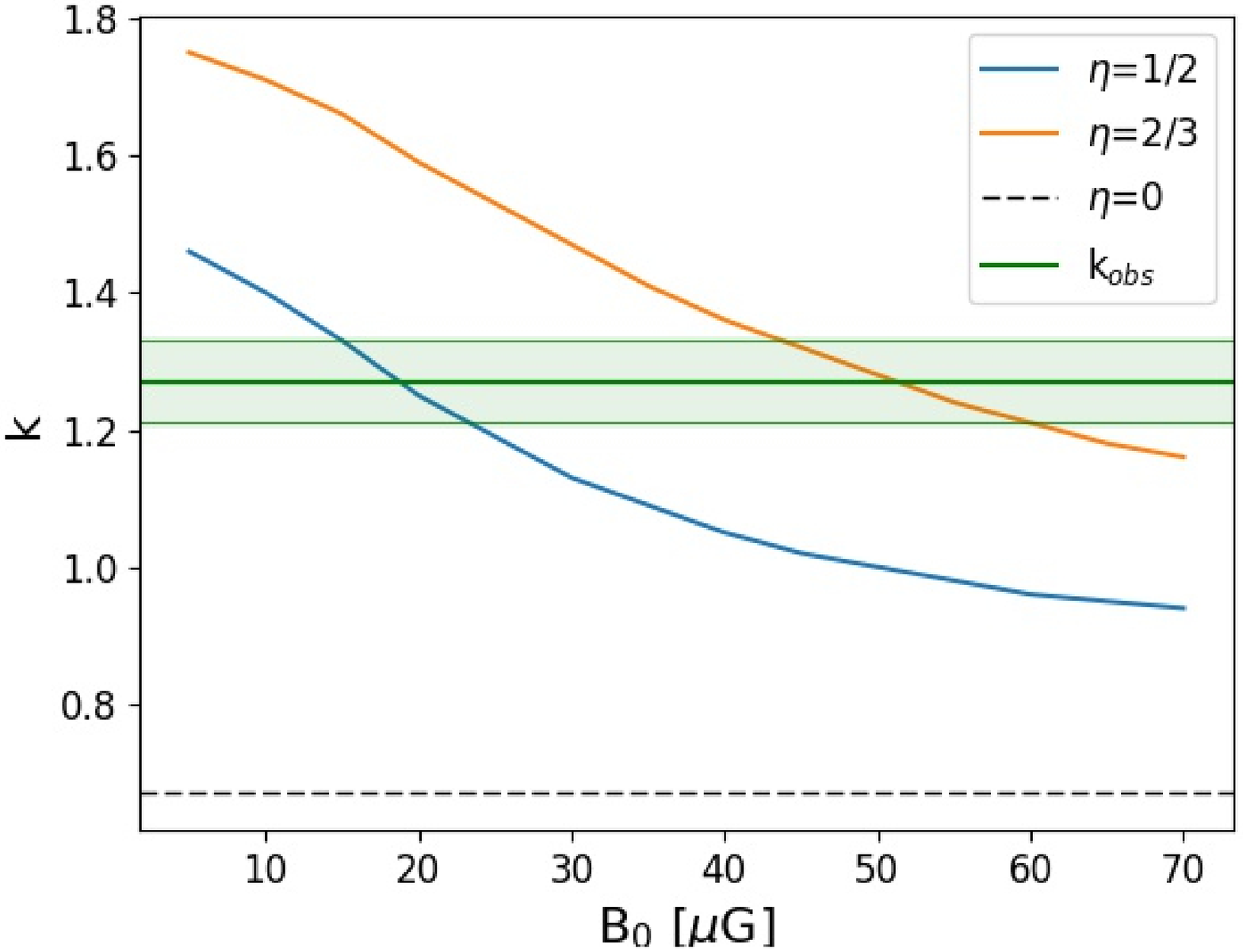}
\endminipage\hfill
\caption{{\it Left:} Results of the Monte Carlo {\it ptp} analysis. The source IDs are reported in Tab. 1 .The horizontal lines point out the average value of the scalings that we measured for MHs (red), the $k$=1 level (black) and the mean value observed for radio halos (green). {\it Right:} Comparison between the observed (green) and the theoretical estimates of $k$ for RBS 797.}
\end{figure}
\vspace{-0.5cm}
\bibliographystyle{aa}
\bibliography{bibliography}
\end{document}